\documentclass[aps,prb,twocolumn,superscriptaddress,floatfix,longbibliography]{revtex4-2}
\usepackage{graphicx}
\usepackage{bm}
\usepackage{wrapfig}
\usepackage{color}
\usepackage{natbib}
\usepackage{hyperref}
\usepackage{amssymb}
\usepackage{amsmath}
\usepackage{geometry}
\usepackage{comment}

\hypersetup{
colorlinks=true,
urlcolor= blue,
citecolor=blue,
linkcolor= blue,
}

\begin{document}

\title{Topological Hall-like behavior of multidomain ferromagnets}
\author{Houssam Sabri}
\email[ ]{houssam.sabri@unh.edu}
\affiliation{Department of Physics and Astronomy, University of New Hampshire, Durham, New Hampshire 03824, USA }
\author{Benjamin E. Carlson}
\affiliation{Department of Physics and Astronomy, University of New Hampshire, Durham, New Hampshire 03824, USA }

\author{Sergey S. Pershoguba}
\affiliation{Department of Physics and Astronomy, University of New Hampshire, Durham, New Hampshire 03824, USA }

\author{Jiadong Zang}
\email[ ]{jiadong.zang@unh.edu}
\affiliation{Department of Physics and Astronomy, University of New Hampshire, Durham, New Hampshire 03824, USA }

\date{\today}

\begin{abstract}
We investigate the emergence of topological Hall-like (THE-like) signals in disordered multidomain ferromagnets. Non-monotonic behavior in Hall resistivity, commonly attributed to topological spin textures such as skyrmions, is produced in a random resistor network model without any chirality. It arises from simple mechanisms of the anomalous Hall effect (AHE) in combination with the domain wall scattering. 
By varying domain configurations and domain wall resistances, we explore the conditions under which the non-monotonic resistivity can be enhanced. Our results emphasize the need for careful analysis in distinguishing between true topological Hall effects and artifacts caused by domain disorders.
\end{abstract}

\maketitle

\section{Introduction}
The topological Hall effect (THE) has emerged as an essential tool for detecting and studying topological spin textures like magnetic skyrmions in chiral magnets. The THE manifests as a contribution to the Hall resistivity that is neither due to the ordinary Hall effect nor the anomalous Hall effect, and often appears as a characteristic hump or dip in the Hall signal as a function of magnetic field \cite{Neubauer2009,Lee2009}. This distinctive nonmonotonic Hall response has been widely interpreted as a hallmark of skyrmion spin textures, which can impart Berry phase effects on conduction electrons \cite{Nagaosa2013}.
The THE arises from the interaction of conduction electrons with the non-coplanar spin texture of skyrmions. As an electron moves through a skyrmion, its spin adiabatically follows the local magnetization direction, accumulating a Berry phase. This Berry phase can be described as an emergent magnetic field experienced by the electron, which deflects its motion and gives rise to the THE \cite{Ye1999,Bruno2004}. The magnitude of the THE is related to the topological charge density of the skyrmion texture.
\begin{figure}
    \centering
    (a)\includegraphics[width=.8\linewidth]{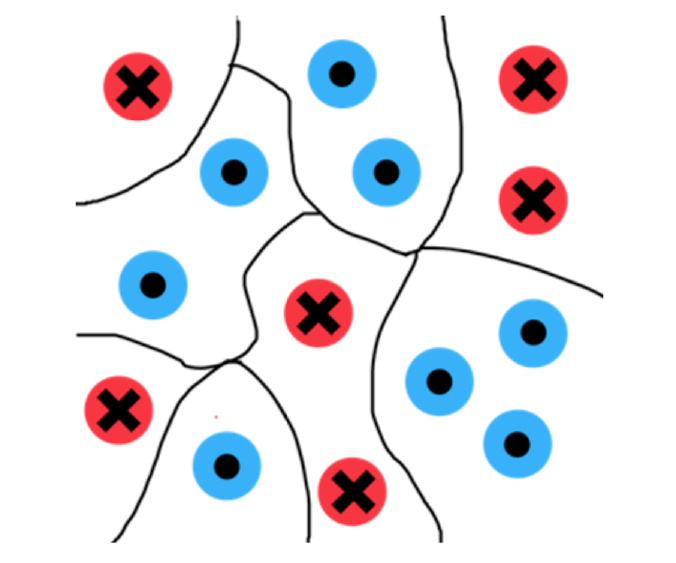} \\
    \vspace{1cm}
    (b)\includegraphics[width=.8\linewidth]{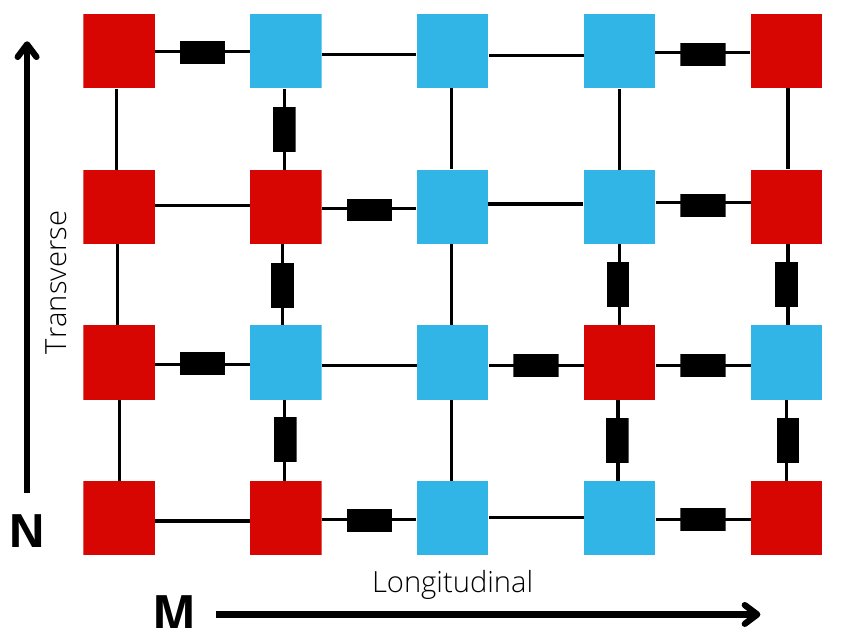} 
     \caption{The blue and red circles in panel (a) depict the opposite spin polarization in different domains of the ferromagnetic material.(a) Ferromagnetic Domains. (b) Resistor Network, where blue boxes indicate a resistor with spin down in the system, and red boxes indicate a spin up resistor; the black boxes depict the giant magneto-resistance effect (domain wall resistance).} 
     \label{fig:domains}
\end{figure}
Early observations of THE in bulk B20 compounds like MnSi, MnGe, and FeGe showed a clear hump feature in the Hall resistivity corresponding to the skyrmion lattice phase \cite{Neubauer2009,Liang2015,Huang2012,Kanazawa2015,Kanazawa2011,Gallagher2017}, which provided strong evidence linking THE to skyrmion formation. Other materials, such as the frustrated triangular-lattice magnet Gd$_2$PdSi$_3$, showed a large THE associated with a field-induced skyrmion phase \cite{Kurumaji2019}. However, the situation is more complex in thin film and multilayer systems that have recently attracted interest for their potential to host room-temperature skyrmions\cite{Woo2016}. In multilayer films composed of ferromagnetic and heavy metal layers, such as [Ir/Fe/Co/Pt] stacks \cite{Soumyanarayanan2017,Raju2019}, oxide heterostructures like SrRuO$_3$/SrIrO$_3$ \cite{Matsuno2016} where electric field control of THE has been demonstrated \cite{Ohuchi2018}, or in (Ca,Ce)MnO$_3$ where a very large THE has been demonstrated \cite{Vistoli2019}, the THE features in these systems are often more subtle than in bulk materials. Careful analysis is required to extract the topological contribution from the measured Hall signal by accounting for the ordinary and anomalous Hall effects. Recent works have revealed several important considerations for interpreting THE in multilayer systems. The magnitude of THE does not always scale simply with skyrmion density, and factors like multi-band transport, interfacial effects, and non-adiabatic processes can complicate this relationship \cite{Maccariello2018,Zeissler2018}. These findings highlight the importance of considering complex spin configurations and magnetic phases beyond isolated skyrmions. 

Furthermore, The effective Hall coefficient relevant to THE may differ from the ordinary Hall coefficient due to multi-band effects, leading to quantitative discrepancies between observed and predicted THE magnitudes and dimensional crossover effects as the film thickness is varied can dramatically influence skyrmion stability and THE \cite{Wang2020}. For example, in Mn-doped Bi$_2$Te$_3$ films, THE was observed only at a critical thickness corresponding to the onset of surface state hybridization \cite{Liu2017} as well as the transitions between isolated skyrmions, skyrmion clusters, and labyrinthine stripe domains, plays a crucial role in determining THE behavior. In these interfacial systems, like the work of work by Raju et al.\cite{Raju2019} on Ir/Fe/Co/Pt multilayers, THE measurements were complemented by direct magnetic imaging techniques like magnetic force microscopy (MFM), which revealed the presence of skyrmions evolving from labyrinth domains, another work by Wang et al. have used in addition to Hall measurement, the polar magneto-optical Kerr effect microscope, and they have observed in multilayer system of Ta/CoFeB/MgO the transformation of an initial labyrinthine domain patter into dense array of skyrmions \cite{Wang2019}, with the same behavior observed in Co/Pt nanostructured films \cite{Sapozhinkov2021}. However, not all studies use imaging techniques to confirm the presence of skyrmions; some of them rely only on the hump in the Hall resistivity measurement as a signature of topological spin textures, like in the Heusler alloy Mn$_2$CoAl where the nucleation and annihilation of skyrmions have been interpreted through THE \cite{Ludbrook2017}. Kumar et al. used the same THE signals as a signature for the presence of antiskyrmions in another Heusler alloys Mn$_{1.4}$Pt$_{1-x}$Pd$_x$Sn and Mn$_{1.4}$Pt$_{1-x}$Rh$_x$Sn \cite{Kumar2020}. Another example is found in the thin films of the Heisenberg ferromagnet EuO, where they interpreted the peaks in Hall resistivity as suggestive of a 2D topologically nontrivial spin structure even in the absence of supporting imaging of this spin textures \cite{Ohuchi2015}.
The interfacial effects in multilayers, such as the coupling between skyrmions across layers and the formation of hybrid chiral domain walls \cite{Legrand2018}, as well as the interface modification of Berry curvature, may lead to the Hall effect peaks and can influence THE in ways not captured by simple models \cite{Groenendijk2020,vanThiel2021}.

While THE remains a powerful probe of skyrmion physics, recent studies have raised important caveats about its interpretation. Gerber pointed out that nonmonotonic Hall signals resembling the THE can arise in heterogeneous ferromagnets with competing ordinary and anomalous Hall effects, necessitating careful analysis in interpreting nonmonotonic Hall signals as definitive evidence for topological spin textures \cite{Gerber2018} and Fijalkowski et al. demonstrated that coexisting magnetic phases in topological insulator heterostructures could produce an AHE-associated response that resembles THE but is not of topological origin \cite{Fijalkowski2020}. They showed that overlapping AHE signals with opposite signs from surface and bulk magnetic phases could mimic THE-like features. These multiples overlapping AHE signals that mimic THE-like signals could also arise in thin films from thickness variation, defect, and interface modification \cite{Kimbell2022}, and SrRuO$_3$ thin films are used as a model system to illustrate these issues. Many studies claiming to observe skyrmions in SrRuO$_3$ can be alternatively explained by AHE inhomogeneities\cite{Wang2020}.
Similarly, Tai et al. developed methods to distinguish genuine THE from artifact "THE" signals from two-component AHE \cite{Tai2022}, where they proposed using minor loop measurements, temperature dependence, and gate dependence to differentiate between actual topological signals and AHE artifacts. To address these concerns, Li et al. reported THE in a simplified topological insulator-magnetic insulator bilayer structure of Bi$_2$Se$_3$/BaFe$_{12}$O$_{19}$ that excludes the possibility of competing AHE signals \cite{Li2021-1}. They observed pure THE at low temperatures (2-3 K) and the coexistence of THE and AHE at intermediate temperatures, with opposite temperature dependencies. Their results demonstrate genuine THE in a topological insulator system \cite{Yasuda2016,Wang2021} and highlight the role of interfacial Dzyaloshinskii-Moriya interaction in stabilizing skyrmions.
These recent developments underscore the need for careful analysis when interpreting THE signals. Direct imaging of spin textures, systematic exclusion of competing effects, and examination of temperature and field dependences are crucial for establishing the topological origin of observed Hall signatures. 

This work aims to prove theoretically that the THE-like signal can be simply induced by the anomalous Hall effect of multidomain ferromagnets. The multidomain structure is modeled by a resistor network. We show that the typical THE-like nonmonotonic behavior in the hall resistivity can also arise from domain wall scattering in the absence of any topological textures.
\section{Plausibility arguement : The Nordheim rule}


Most magnetic multilayers exhibiting THE-like signal have large perpendicular magnetic anisotropy (PMA). As a result, labyrinth domains, as mixture states of spin-up and -down magnetic domains, are emergent at the magnetization reversal[see Fig.~\ref{fig:domains}(a)]. These structures resemble binary alloys, which consist of two elements. In these binary alloys, the Nordheim rule describes the longitudinal resistivity \cite{Nordheim1931}. In this case, the resistivity of the system is proportional to the product of the concentrations of each element, i.e.,
\begin{equation}
\rho_{xx}=\rho_0+\rho_1 x (1-x)
\end{equation}
where $\rho_0$ is the resistivity of both elements, assuming they are the same. $x$ and $1-x$ are concentrations of each element \textcolor{black}{and $\rho_1$ is the residual resistivity due to impurities and is proportional to the concentration of each element}. In the dilute limits, $x\rightarrow0$ and $x\rightarrow1$, $\rho_{xx}$ is proportional to the impurity concentration, consistent with the transport theory. The resistivity is thus maximized at $x=1/2$ where the most disorder occurs, i.e., when there is an even mix of the components. Applying the Nordheim rule to the multidomain system under investigation, $x_\uparrow=(1+s)/2$ and $x_\downarrow=(1-s)/2$ are percentages of up and down domains, respectively. $s=M/M_s$ is the normalized magnetization , where $M$ is the total magnetization and $M_s$ is the saturation magnetization. The resulting resistivity is thus quadratic with respect to $s$.
\begin{equation}
    \rho_{xx}=\beta-\gamma s^2. \label{nordheim}
\end{equation}
Here, $\beta$ and $\gamma$ are some phenomenological parameters,  and normalized magnetization $1\ge s \ge -1$ governs the degree of disorder in the material. Notice that the longitudinal resistivity reaches a maximum at $s=0$, which implies an even amount of domains. It is also conceivable that the transverse conductivity scales linearly with average polarization (e.g., by virtue of the anomalous Hall effect) $\sigma_{xy}=\alpha s $,
combining these equations yields the Hall resistivity in a disordered binary ferromagnet.
\begin{equation}
  \rho_{xy}\approx \sigma_{xy}\, \rho^2_{xx} = \alpha s (\beta-\gamma s^2)^2 \label{hall_resist_nordheim}
\end{equation}
the behavior of the Hall resistivity has a clear deviation from the linear anomalous Hall resistivity curve $\rho^A_{xy}=\alpha(\beta-\gamma)s$ connecting two ends $s=\pm1$ \textcolor{black}{\cite{Nagaosa2010}}. The ordinary Hall effect is not included since orbital coupling to the external magnetic field is not considered here. The discrepancy between Eq.(\ref{hall_resist_nordheim}) and the anomalous Hall is what researchers often label as the topological Hall effect discussed previously. Therefore, the topological Hall-like signal can be, in principle, emergent in these simple systems as well. The Hall resistivity in Eq.(\ref{hall_resist_nordheim}) can also exhibit non-monotonic behavior. The equation implies $\rho_{xy}(s)$ may have maxima at $s^2=\frac{\beta}{5\gamma}$. But again, it is worth emphasizing that one does not need non-monotonicity to claim the presence of a topological Hall-like effect; the nonlinear deviation from the anomalous Hall effect is sufficient.


\section{Model approach. Random resistor network.}
In order to support the hand-waving argument from the previous section, a resistor network model is introduced. A two-dimensional ferromagnetic material schematically presented in Fig.~\ref{fig:domains}(a) is investigated. It is composed of domains that are described by an out-of-plane local polarization. We assume that longitudinal conductivities, denoted here as $G_1$ and $G_2$, are identical for domains with distinct polarization. In contrast, opposite domains have opposite signs of the anomalous Hall conductivity $G_H$ defined by their local polarization. We exclude the conventional Hall effect from the start. In addition, the scattering of electrons moving across domain boundaries is described by a longitudinal resistance $R$ following the domain wall resistance. 

To account for all these effects, we employ a discrete random resistor network (RRN) illustrated Fig.~\ref{fig:domains}(b) inspired by Ref.~\cite{parish2003}. Below, we discuss the details of implementation.  

\subsection{Resistor Network}
\begin{figure}[h]
\centering
  \includegraphics[width=\linewidth]{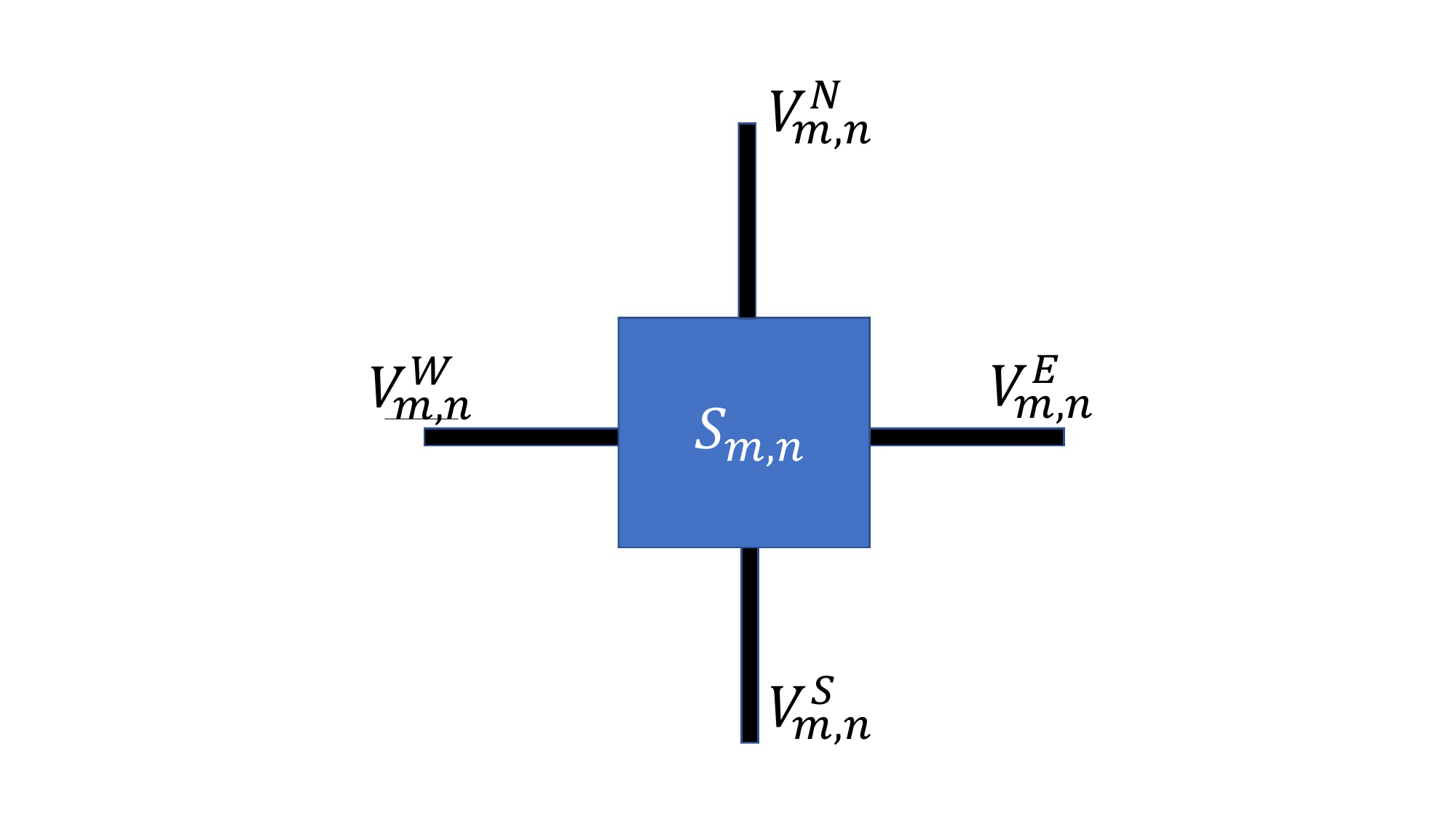}
  \caption{Single unit from the resistor network. The positive sign current is flowing into the resistor from west (W) to east (E) or from south (S) to north (N) in the transverse direction. The west end of the lattice was set to 1 unit of voltage, and the east end was set to 0. }
  \label{fig:unit_res}
\end{figure}
The network is shown in Fig.~\ref{fig:domains}(b). It consists of four-terminal (square) and two-terminal (square) resistors. The former describes current flow within domains, whereas the latter represents a possible added resistance to domain-wall scattering between two opposite domains. Ultimately, the resistor network could be any M-by-N size. The periodic boundary condition is applied in the transverse direction so that the bottom and top of the resistor network are connected into a sort of cylinder. In this way, the current would always be driven west to east in some way, although the path it would take would vary depending on conditions imposed by the Hall effect and the domain wall resistance. The resistor network is considered to have discrete conductivity and spin-dependent.

\subsection{Electric Current in a Single Unit}

For our purpose, the current was always defined as going into the unit and having four paths. The current for each terminal can be defined using Ohm's law in terms of conductivity $I=VG$, refer to Fig.~\ref{fig:unit_res}:
\begin{widetext}
\begin{equation}
\begin{aligned}
    & I^{m,n}_E=G_1(V_{m,n}^E-V_{m,n}^W)+G_2(V_{m,n}^E-V_{m,n}^N)+G_2(V_{m,n}^E-V_{m,n}^S)+G_HS_{m,n}(V_{m,n}^N-V_{m,n}^S)\\
    & I^{m,n}_N=G_2(V_{m,n}^N-V_{m,n}^S)+G_1(V_{m,n}^N-V_{m,n}^E)+G_1(V_{m,n}^N-V_{m,n}^W)+G_HS_{m,n}(V_{m,n}^W-V_{m,n}^E)\\
    & I^{m,n}_S=G_2(V_{m,n}^S-V_{m,n}^N)+G_1(V_{m,n}^S-V_{m,n}^E)+G_1(V_{m,n}^S-V_{m,n}^W)+G_HS_{m,n}(V_{m,n}^E-V_{m,n}^W)\\
    & I^{m,n}_W=G_1(V_{m,n}^W-V_{m,n}^E)+G_2(V_{m,n}^W-V_{m,n}^N)+G_2(V_{m,n}^W-V_{m,n}^S)+G_HS_{m,n}(V_{m,n}^N-V_{m,n}^S)
\end{aligned}
\label{unit_eqs}
\end{equation}
\end{widetext}
$G_1$ and $G_2$ are the conductances of the system along the x and y directions, respectively. $G_H$ is the term that controls the conductance strength of the Hall effect for our system. The indices $m$ and $n$ loop over the entire M-by-N system \textcolor{black}{and $S_{m,n}$ is a random variable for the spin polarization of the resistor modeling a single domain}. For each resistor in the system shown in Eq.~(\ref{unit_eqs}), four equations are defined as the voltage drop across each junction on the resistor. The additional term, depending on the unit's spin, accounts for the Hall effect.  
We solve Kirchhoff's laws, ensuring that the total current into each resistor vanishes, which transforms our system into a set of linear equations that are solved using the Jacobi iterations. 


\subsection{Generating the spin configuration} 
To model disordered ferromagnetic domains in our resistor network, we randomly assign a spin variable to each resistor in the system in the form of a +1 or -1, the probability of which was controlled by the average magnetization $s$. This means that the modeled ferromagnet's average magnetization will range between 1 and -1, with $s=1$ and $s=-1$ corresponding to a magnet being fully polarized in the opposite directions. Then, we define the spin $S_{m,n}$ of a given unit using the equation 
\begin{equation} 
  S_{m,n} = \left\{\begin{array}{cc}
      1, & {\rm if\,}\textcolor{black}{p}<(s+1)/2, \\
      -1, & {\rm if\,}\textcolor{black}{p}>(s+1)/2,
  \end{array}\right. 
  \label{SpinEQ}
\end{equation} 
where \textcolor{black}{$p$} is a randomly generated number in the interval $\textcolor{black}{p} \in (0,1)$. By design, the expectation value of the random variable $\langle S_{m,n}\rangle = s$ is related to the average polarization $s$.



\subsection{Domain wall scattering induced resistance} 
Domain-wall scattering-induced resistance is another spin-induced effect that would dictate how the current moves through our random resistor network. This is manifested as an additional resistor between each junction in the lattice with two current inputs. If the adjacent resistors in the system had the same spin, then there is no domain wall and, thus, no resistance. However, if the neighboring resistor had opposite spins, a resistance factor would be added to the current traveling through that junction, which, in turn, leads to the current tending to travel through domains of the same spin resistors. The effect was modeled as an additional term governing the currents in the form of
\begin{equation}
R_{m,n}^{m+1,n}=R\cdot(1-S_{m,n}\cdot S_{m+1,n})
\end{equation}
for the current in the x-direction where $R$ is the strength of the domain wall resistance for each junction. $S_{m,n}$ and $S_{m+1,n}$ denote the adjacent spins the current traveling through the junction is associated with m, and n are the indices for the generalized system. Naturally, there is also a domain wall resistance for the y-direction as well: 
\begin{equation}
R_{m,n}^{m,n+1}=R\cdot(1-S_{m,n}\cdot S_{m,n+1}).  
\end{equation}

\subsection{Computation of longitudinal and transverse conductivities} 

By varying the average overall magnetization, $s$,  and averaging over current and voltage for the system, the average Hall conductivity was acquired for each value of $s$. The average of the longitudinal and transverse conductivity and resistivities were determined by multiple realizations for each $s$ (50 realizations for our results). Thus, the term we evaluated over the system is
\begin{equation}
    \sigma_H=\sigma_{xy}=\frac{\bar I_y}{\bar V_x}
\end{equation} 
where $\bar I_y$ is the current averaged over the system in the transverse direction and $\bar V_y$ is the average voltage in the x direction. Likewise, we can acquire longitudinal conductivity: 
\begin{equation}
    \sigma_L=\sigma_{xy}=\frac{\bar I_x}{\bar V_x}
\end{equation} 
With $\bar I_x$ is the average current in the x direction, and using these equations, we determine the longitudinal and Hall resistivity values for the system: 
\begin{align}
\rho_H=\rho_{xy}=\frac{\sigma_H}{\sigma_H^2+\sigma_L^2}, \,\,\,\,\ \rho_L=\rho_{xx}=\frac{\sigma_L}{\sigma_H^2+\sigma_L^2}.
\end{align}
  
\begin{figure*}
    \centering
    (a)\includegraphics[width=.3\linewidth]{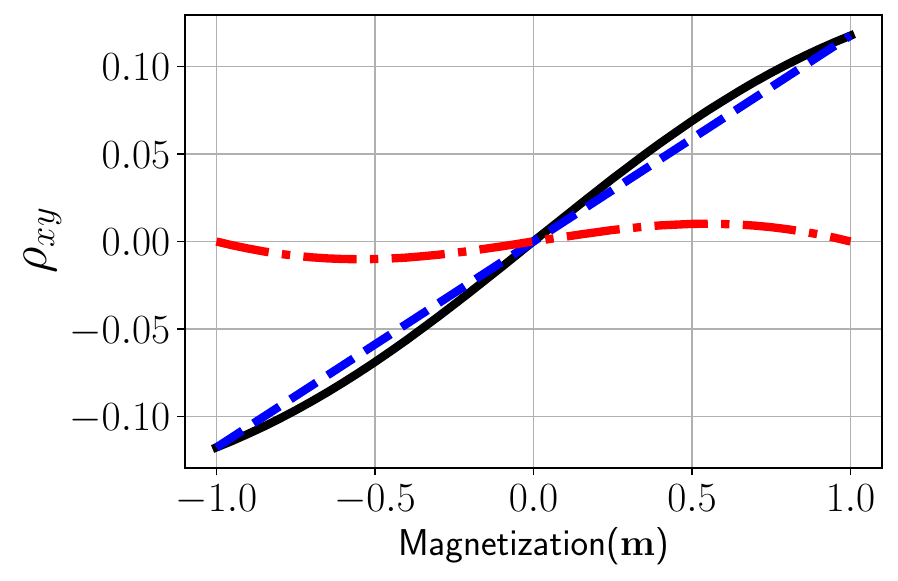}
    (b)\includegraphics[width=.3\linewidth]{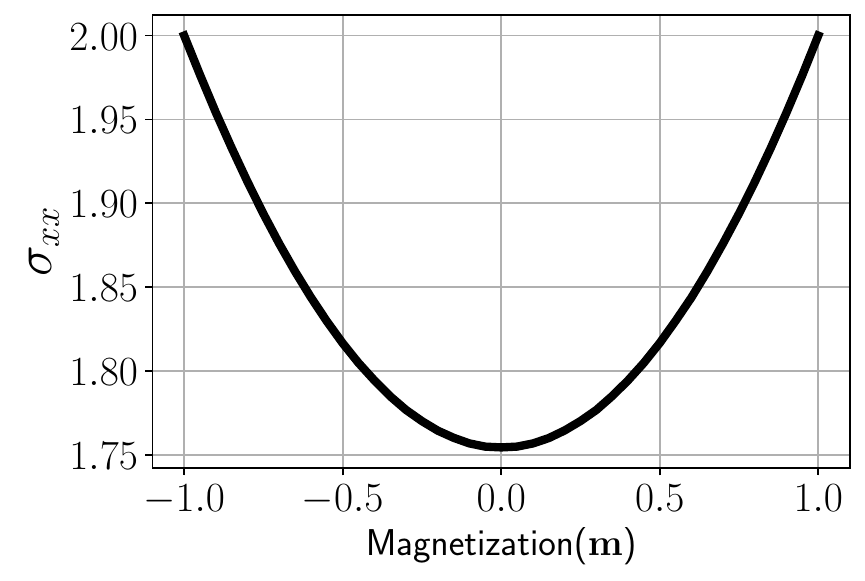}
    (c)\includegraphics[width=.3\linewidth]{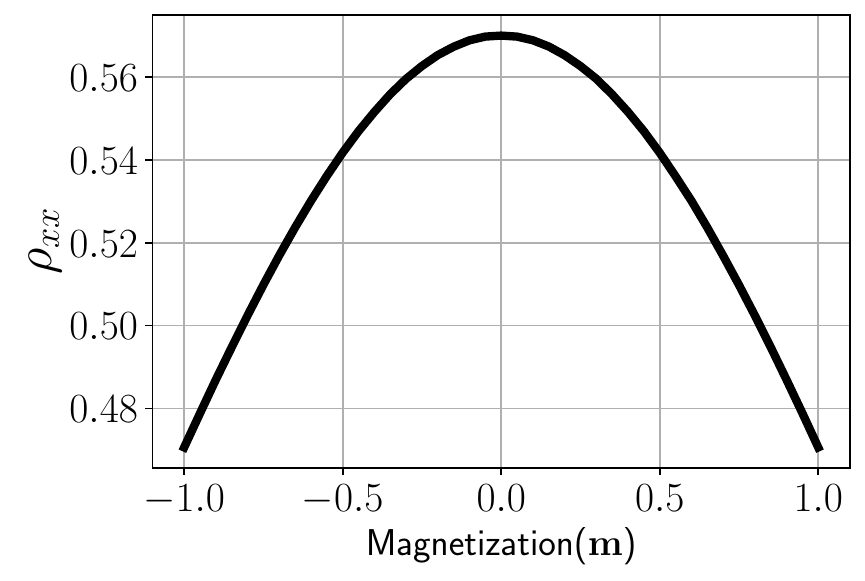}  \\
    (d)\includegraphics[width=.3\linewidth]{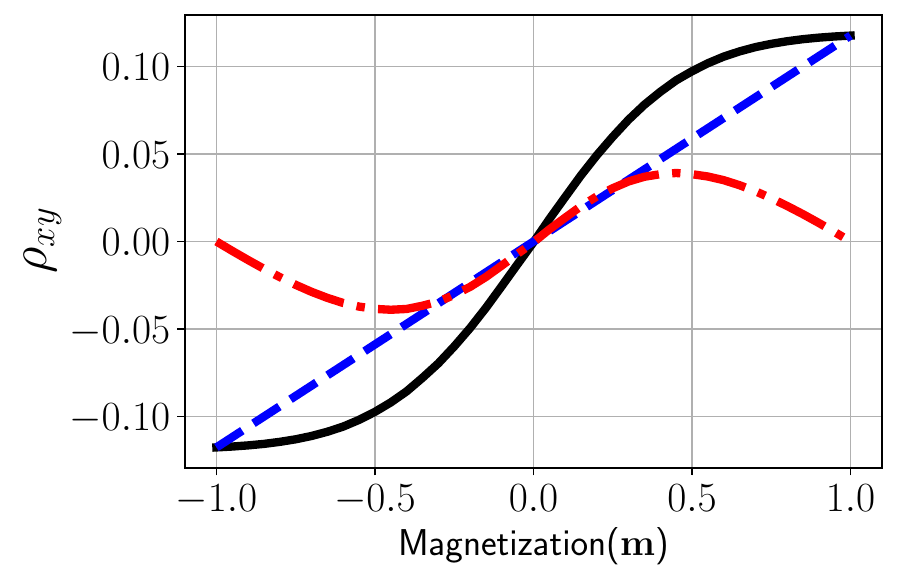} 
    (e)\includegraphics[width=.3\linewidth]{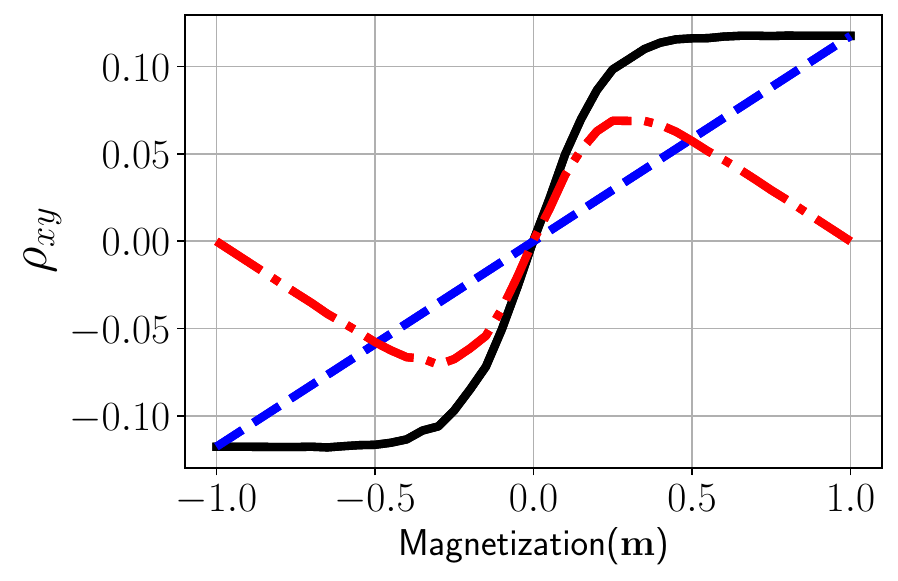} 
    (f)\includegraphics[width=0.3\linewidth,height=0.25\linewidth]{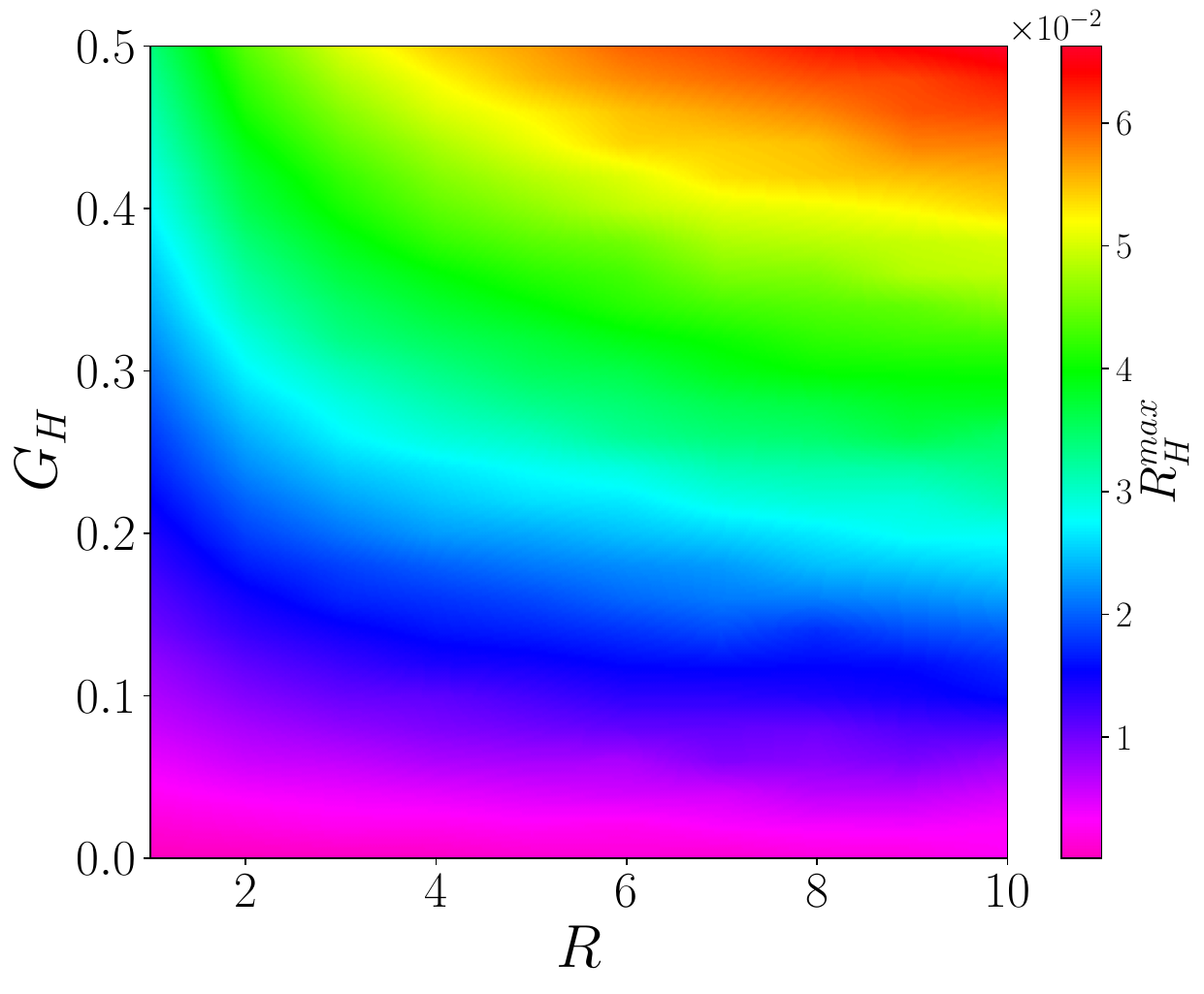} 

     \caption{Hall resistivity (black solid line) and conductivity versus normalized magnetization of a ferromagnet. Decomposed in two contributions: a linear in normalized magnetization, shown with blue dashed lines, and the remainder, shown with the red dash-dotted curves. The graphs are plotted for the following parameters $G_1 = G_2 = 1$, $G_H = 0.5$ and with different domain wall resistances for $R =  0.1$ in (a), with its corresponding longitudinal conductivity (b) and longitudinal resistivity (c) and the same Hall decomposition for  $R = 1$ in (d), and $R = 10$ in (e).\textcolor{black}{ The subfigure (f) shows the RHR maximum $R^{max}_H$} as function of Hall conductivity of the single unit ($G_H=0-0.5$) and domain wall resistance ($R=0-10$). } 
     \label{fig:results_for_large_system}
\end{figure*}

\section{Results and discussion} \label{sec:results}

Jacobi iteration is used to solve Kirchhoff's equation and calculate the longitudinal and Hall conductivity as a function of the magnetization. The resistivity is computed by the inverse of the conductivity tensor, and the Hall resistivity can be identified.  
The results for a 500-by-500 system are presented in Fig.~\ref{fig:results_for_large_system}. The Fig.~\ref{fig:results_for_large_system}(b, c) show the variation of longitudinal resistivity and conductivity, respectively, as a function of magnetization $m=M/M_s$, which is normalized to the saturation magnetization. The domain wall resistance $R =  0.1$. The curve of the longitudinal resistivity does obey Nordheim's rule. 
The resulting Hall resistivity, represented in the black solid line of Fig.~\ref{fig:results_for_large_system}(a), clearly deviates from linearity. The experimental determination of the anomalous Hall effect, $\rho^{xy}\propto m$,  is indicated by the straight dashed line connecting two fully polarized states. The Residual Hall Resistivity (RHR), defined as the subtraction of the anomalous Hall resistivity from the total Hall resistivity, shows a non-monotonous behavior. Since orbital effects are not considered here, the ordinary Hall effect is not included. The RHR, shown in the dashed-doted line, is usually interpreted as THE. 
However, the non-monotonicity of RHR does not originate from a topological origin in this collinear spin model. All we considered is a binary mixture of spin-up and spin-down domains that mimic disordered ferromagnets that exhibit labyrinth-like behavior domains. There is no room for topological spin textures to exist, and therefore, any interpretation of the non-monotonous RHR arising from topological origin is erroneous. The emergence of the RHR similar to THE here solely depends on how we separate the anomalous Hall effect from the total Hall resistivity.  

All we considered here is the anomalous Hall effect only. This random resistor network model serves as a prototype model to show that a peak in the RHR alone is not a sufficient argument for the existence of topological spin textures and should not always be labeled as a 'THE'. The assumption that AHE is always linear with magnetization is often not valid since there is always a non-linear part labeled as the higher order anomalous Hall effect which is non-negligible in systems that exhibit large Berry curvature in momentum space \cite{SodemannFu2015}, like in Weyl semi-metals \cite{Li2021,Cao2022,Xiao2024} and transition metal dichalcogenides \cite{Kang2019}.
\begin{figure*}
    \centering
    (a)\includegraphics[width=.3\linewidth]{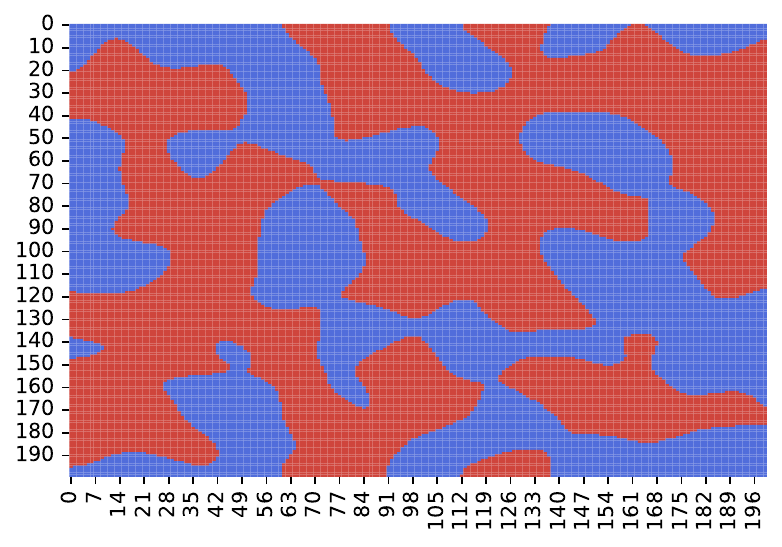} 
    (b)\includegraphics[width=.3\linewidth]{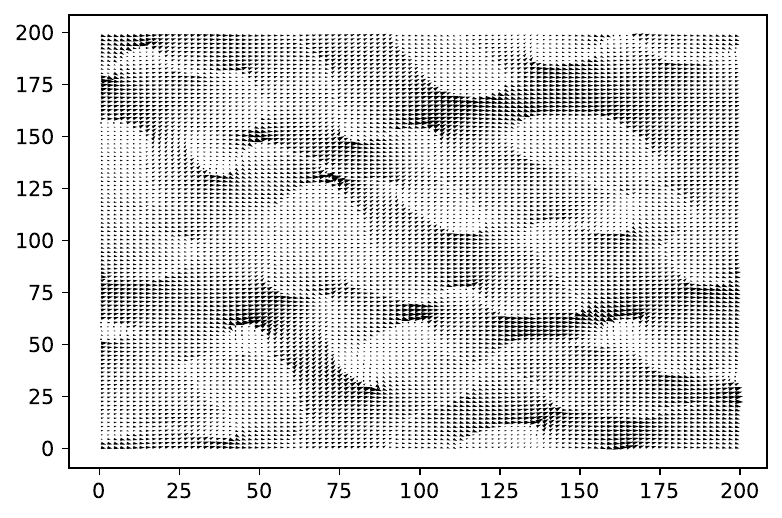}
    (c)\includegraphics[width=.3\linewidth]{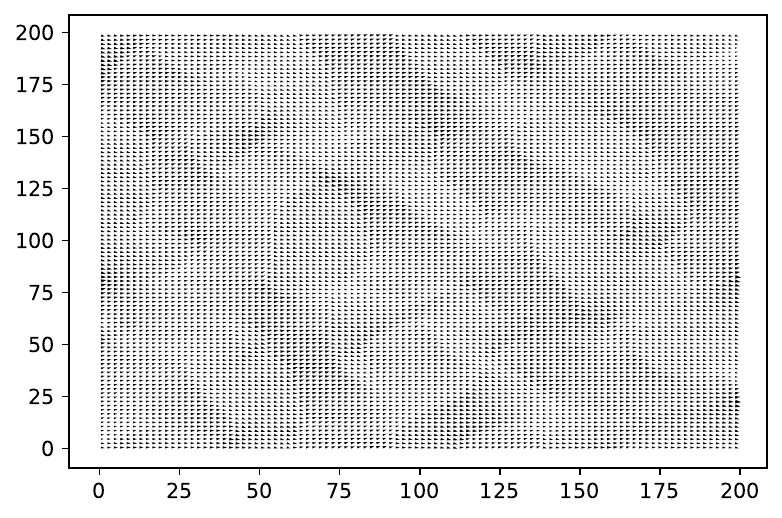}\\
   (d)\includegraphics[width=.3\linewidth]{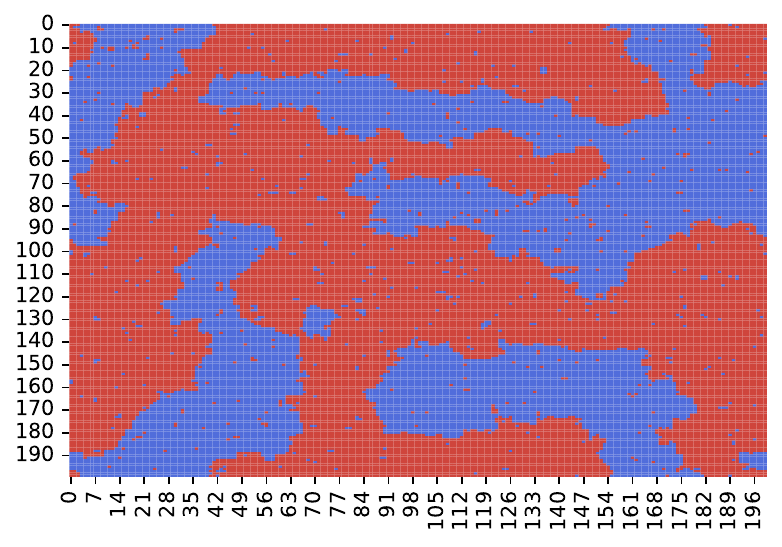} 
    (e)\includegraphics[width=.3\linewidth]{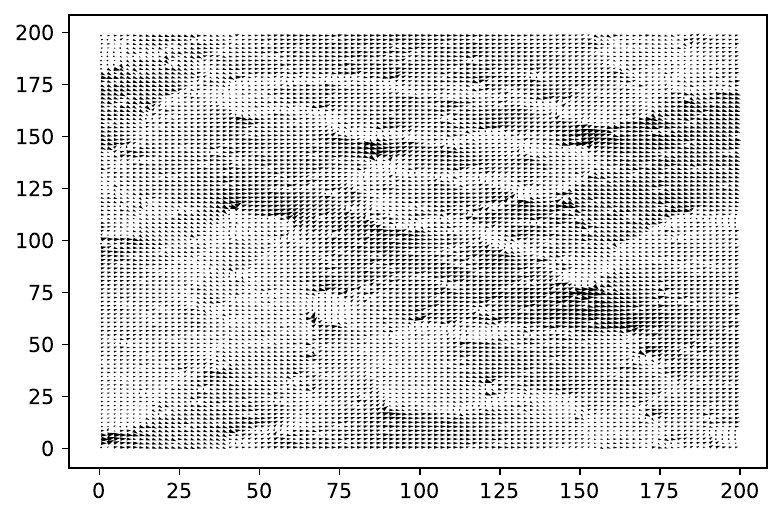}
    (f)\includegraphics[width=.3\linewidth]{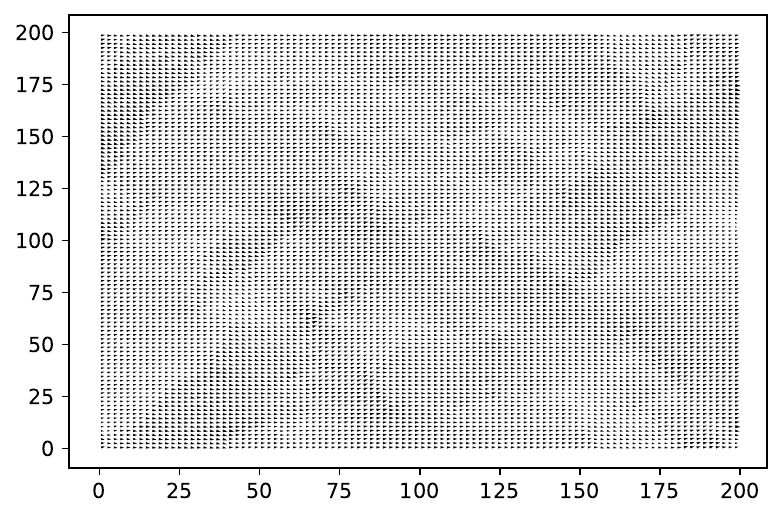}
     \caption{Current profile for different configurations and domain wall resistances: Results of modeling in a finite  200-by-200 sample with 60/40 (up/down) domains randomly generated for parameters $G1 = 1$, $G2 = 1$, $GH = 0.5$ and (a) its corresponding spin domains configuration. (b) The current profile of the randomly generated configuration exhibits a labyrinth-like pattern avoiding domain walls for  $R = 10$, and (c) the current profile of the same configuration for $R = 0.1$ shows the easy scattering between domains. Result for  200-by-200 sample for (60 spin up/40 spin down) obtained by Ising Monte Carlo simulations with the same previous parameters $G1 = 1$, $G2 = 1$, $GH = 0.5$ with (d) the Ising model Spin domains configuration, (e) its corresponding current profile with same behavior for  $R = 10$ and (f) the current profile for $R = 0.1$ } 
     \label{fig:results_for_small_system1}
\end{figure*}

Furthermore, Fig.~\ref{fig:results_for_large_system}(d-e) show the evolution of the RHR as the domain wall resistance increases, and Fig.~\ref{fig:results_for_large_system}(f) shows the contour plot of the maximal RHR $R^{max}_H$ as a function of both the Hall conductance $G_H$ and $R$. It is not surprising to see that the non-monotonicity of RHR is enhanced with the increase of Hall conductance since the whole transverse resistivity scales with $G_H$. However, it is interesting that as the domain wall resistance increases, the total hall resistivity nonlinear behavior is more pronounced.

To understand the relation between RHR and the domain wall resistance $R$, the current distribution is studied, as shown in Fig.~\ref{fig:results_for_small_system1}. At positive magnetization, small islands of down spins are formed (see Fig.~\ref{fig:results_for_small_system1}(a)). As shown in Fig.~\ref{fig:results_for_small_system1}(b), due to the large domain wall resistance, current flowing across opposite domains is suppressed so that a large portion of current stays in the \textcolor{black}{same spin domains}. \textcolor{black}{Since the disorder is applied onsite in this model and the correlation between neighbors is fully ignored, we have also conducted Monte Carlo simulations of the Ising model to reveal the current distribution of a more realistic scenario}. The Hamiltonian is simply
\begin{equation}
H = -J\sum_{\langle(m,n),(l,k) \rangle} S_{m,n} S_{l,k} - h \sum_{m,n} S_{m,n}
\end{equation}
with $J$ the exchange coupling and $h$ the applied field. A 200-by-200 sample thermalized at $h=0.01J$ and $T=1.9J$ is prepared, whose snapshot is shown in Fig.~\ref{fig:results_for_small_system1}(d). Due to the coupling between neighboring spins in the Monte Carlo simulation, domains of opposite spins are formed. Using the same parameters $G_1 = 1$, $G_2 = 1$, $G_H = 0.5$ and $R = 10$ , the current distributions are shown in Fig.~\ref{fig:results_for_small_system1}(e). Avoided cross-domain scattering is more obvious. Mostly, current stays in the majority domain. 

As a result of the uneven distribution of current, the Hall resistivity is not an average of the spin population but is dominated mainly by the majority spin. The deviation between the Hall resistivity and overall magnetization thus emerges. Actually, when the domain wall resistance becomes larger, the longitudinal resistivity shown in Fig.~\ref{fig:results_for_large_system} deviates from Nordheim's rule in proximity to the fully polarized state. The reason for such deviation is the same here. 
\begin{figure}
    \centering
    (a)\includegraphics[width=.71\linewidth]{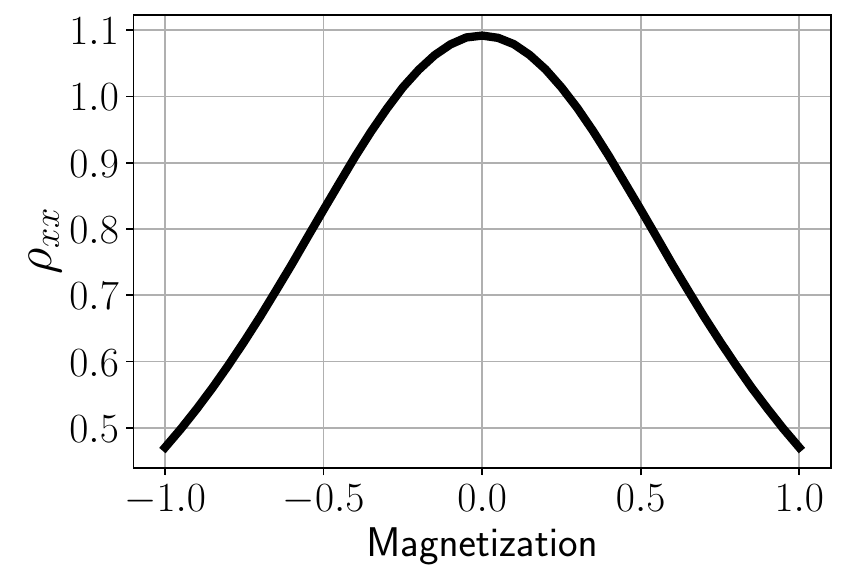} 
    (b)\includegraphics[width=.71\linewidth]{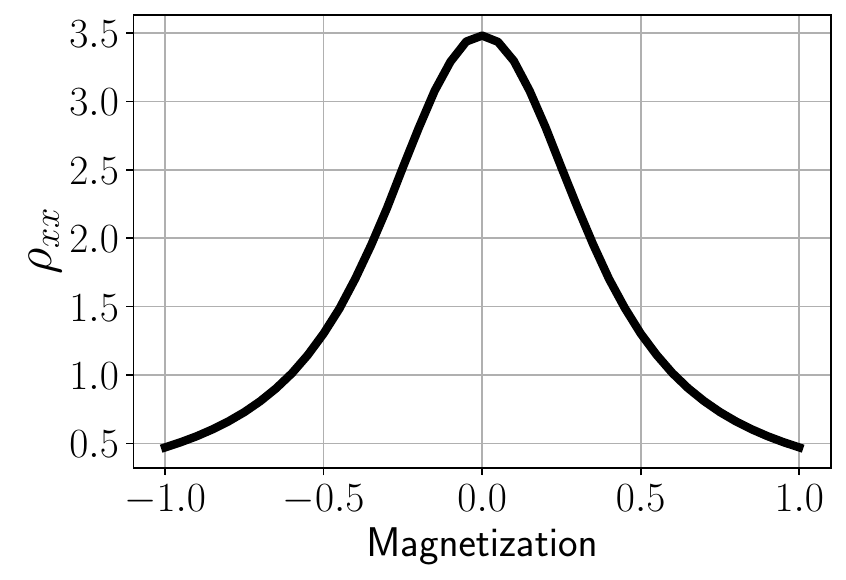}
     \caption{\textcolor{black}{Longitudinal resistivity as a function of average magnetization of system for domain wall resistance (a) $R=1$ and (b) $R=10$ showing the deviation from the Nordheim parabolic behavior as $R$ increases.} } 
     \label{fig:deviationNordheim}
\end{figure}

\begin{figure*}
    \centering
    (a)\includegraphics[width=.3\linewidth]{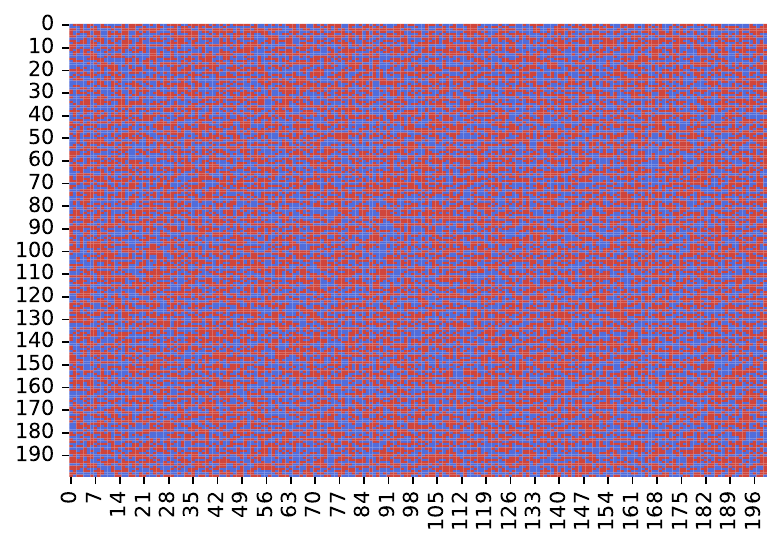} 
    (b)\includegraphics[width=.3\linewidth]{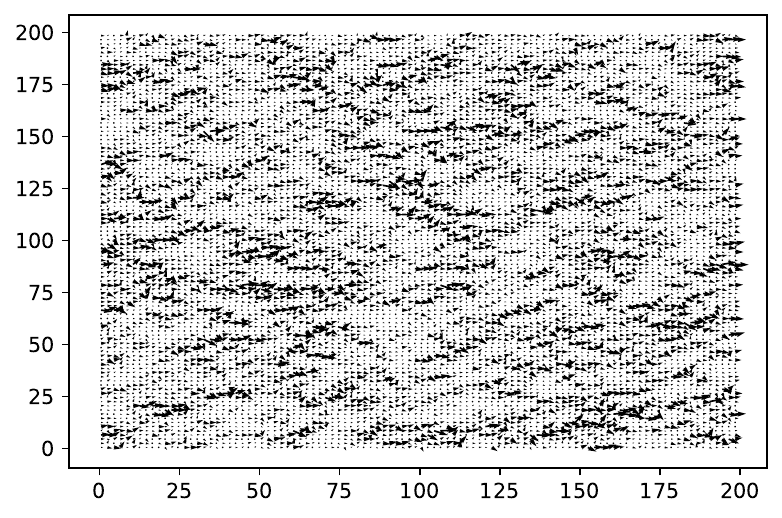}
    (c)\includegraphics[width=.3\linewidth]{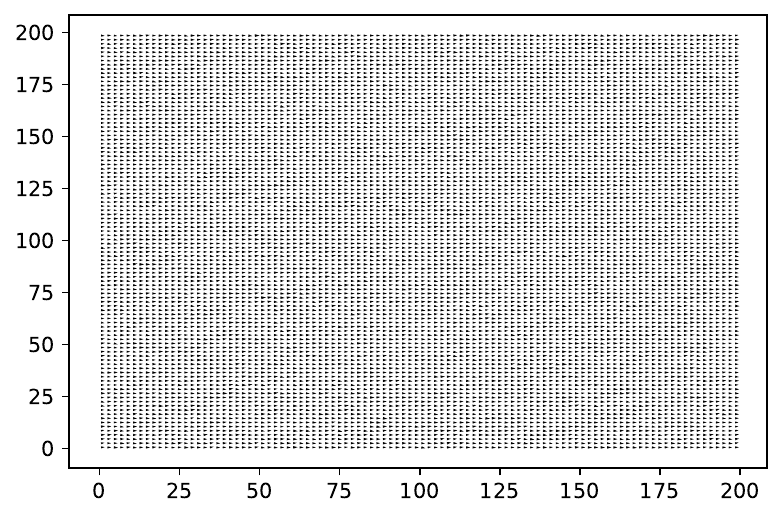}\\
   (d)\includegraphics[width=.3\linewidth]{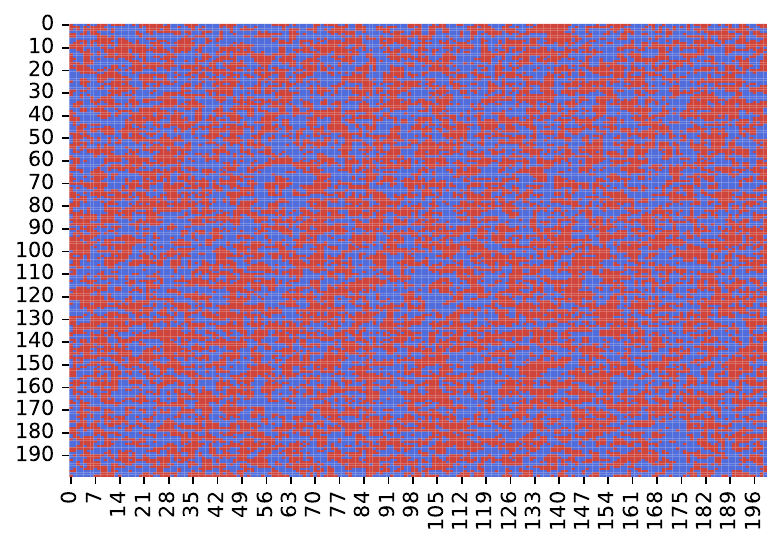} 
    (e)\includegraphics[width=.3\linewidth]{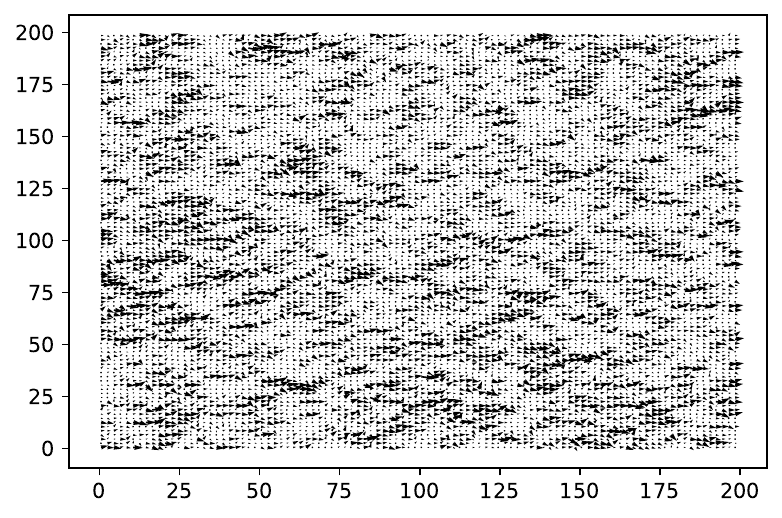}
    (f)\includegraphics[width=.3\linewidth]{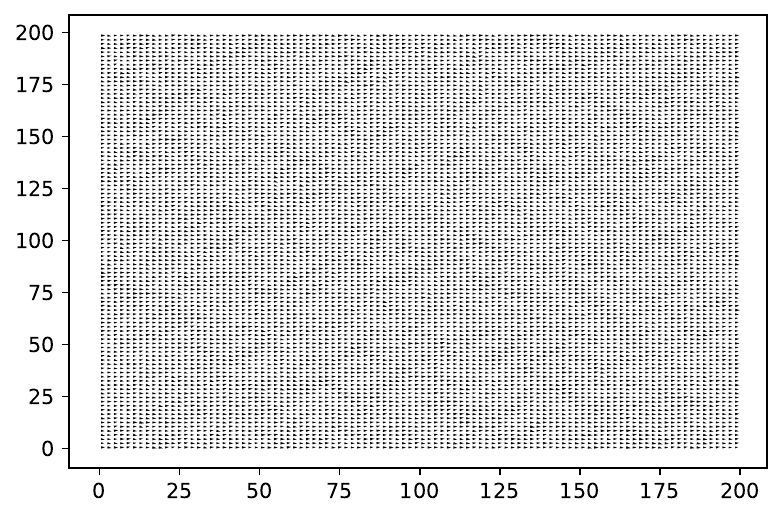}

     \caption{(a)Results of randomly generated spin domain with 50/50 split (maximum disorder) of spin channels with a current profile for (b) $R=10$ and (c)  $R=0.1$. Result for (50/50 split) obtained by Ising Monte Carlo simulations with the same previous parameters $G1 = 1$, $G2 = 1$, $GH = 0.5$ with (d) the Ising model Spin domains configuration, (e) its corresponding current profile with same behavior for  $R = 10$ and (f) the current profile for $R = 0.1$} 
     \label{fig:results_for_small_system2}
\end{figure*}
Once the domain wall resistivity is reduced to $R=0.1$, as shown in Fig.~\ref{fig:results_for_small_system1} (c) and (f), current is more easily scattered into minority domains. The Hall resistivity is, therefore, closer to the average population of up and down domains and becomes closer to anomalous Hall resistivity. RHR is thus suppressed. On the other hand, when the magnetization becomes close to zero, i.e., a completely disordered state with an equal population of up and down domains, as shown in Fig.~\ref{fig:results_for_small_system2}, the current becomes evenly populated among two domains and the Hall resistivity becomes zero as well. This recovers the anomalous Hall effect. The deviation between total Hall resistivity and anomalous Hall resistivity is maximized at a moderately magnetized state.


In this work, we developed a theoretical framework to explain the emergence of a topological Hall effect-like signal in multidomain ferromagnetic systems. Through the use of a random resistor network model, we show that non-monotonic Hall resistivity behavior — often interpreted as evidence of topological spin textures such as skyrmions — can arise from much simpler mechanisms such as domain wall scattering and the anomalous Hall effect. Our model successfully captures how a disordered arrangement of ferromagnetic domains, each with its polarization, generates a Hall signal that mimics THE without necessitating the presence of any topologically nontrivial structures.

\textcolor{black}{Our domain-scattering-based RRN model typically describes a single ferromagnetic layer or thin film in which electrons experience local, up and down magnetization domains (with a conventional anomalous Hall coefficient of one sign overall) and resistivity at domain walls, where conduction is partially impeded or scattered. In the two components AHE there are two magnetic channels with opposite signs of anomalous Hall coefficients and different coercivity fields, in which the hump emerges in between two coercivity fields once one channel is reversed while the other is not. A key feature of this mechanism is the presence of steps of the M-H curve, and the hump is located at the negative side of the external field. However, our mechanism does not require two magnetic channels. Only one species of magnetic system is involved. The hump is located on both the positive and negative sides of the external field as long as disordered domains are formed. The nonmonotonic Hall response emerges from the spatial arrangement of these up/down domains and their boundaries. As the external magnetic field changes, the fraction and shape of up/down domains evolve, altering where current can flow easily and is forced across domain walls. This can create extra humps or dips in Hall resistivity even without any sign change in the local AHE.
Empirically, both scenarios may produce hump-like or nonmonotonic features in the Hall effect, potentially mimicking a “topological Hall” signal. However, distinguishing them involves using imaging techniques like MOKE microscopy to see whether the material is genuinely single-phase with up/down domains or has multiple regions, each displaying different magnetization loops, and Field-dependent and temperature-dependent measurement to check whether the nonmonotonic feature lines up with domain expansion or separate phase reversals.} 

\textcolor{black}{Our RRN approach demonstrates how ordinary domain inhomogeneity and domain-wall scattering can, on their own, produce Hall loops with hump-like features often misattributed to a real-space Berry phase. This mechanism is especially relevant in thin-film ferromagnets with strong perpendicular magnetic anisotropy (strong uniaxial anisotropy). The model assumes purely out-of-plane magnetization; any domain wall is modeled as an additional resistance. The model typically uses the same single-sign anomalous Hall coefficient per domain. Still, one might need a multi-component approach for more accurate results in materials exhibiting multi-band transport or doping inhomogeneities.
Our RRN model applies broadly to thin-film ferromagnets where measurable domain walls form upon up-and-down domain formation. For example, in thin film SrRuO$_3$ (SRO), for instance, cooling below Tc forms stripe domains that raise the total resistivity by about 0.46 $\mu\Omega.cm$, which is an increase of around 11\% in longitudinal resistivity compared to monodomain case (at high fields)\cite{Klein2000}. This is comparable in the RRN model to the case where domain wall resistivity is around R=0.1 leading to 18\% increase in longitudinal resistivity as seen in Fig. \ref{fig:results_for_large_system}.(c).
In ultrathin manganite films like Pr$_{2/3}$Sr$_{1/3}$MnO$_3$ (PSMO) on LaAlO$_3$ substrates with compressive strain, strong PMA, and labyrinth domains have been observed\cite{Wu1999}. The labyrinth domain state is associated with a giant magnetoresistance, where the domain walls increase the resistance up to 90\% compared to the saturated monodomain state. For our model, this is the case where we have a domain wall resistance R=1. In such a case, we expect the maximum RHR to be around 30\% of the total Hall resistivity at saturation. The same behavior is also seen in thin film La$_{0.7}$Sr$_{0.3}$MnO$_3$ (LSMO); however, this effect is much smaller in LSMO films compared to PSMO\cite{Li2001}.
Non-oxide materials show similar behavior in a variety of samples, for example, FePd thin films \cite{Danneau2002}. However, the effect of the wall magnetoresistance is much smaller and is around 10\% therein, which puts it in the same order of magnitude as SRO.} 

This study provides a crucial clarification in the ongoing exploration of THE in magnetic systems. While skyrmions and other topological spin textures are known to produce distinct Hall signatures, our results demonstrate that similar features can emerge in systems where no such textures exist. This suggests that caution is necessary when interpreting experimental data, as the mere presence of a hump or dip in Hall resistivity is not a sufficient criterion for identifying topological phenomena. Factors such as domain structure, domain wall scattering, and material inhomogeneities can contribute significantly to these observations. Furthermore, our simulations revealed that the strength of the domain wall resistance plays a key role in enhancing the non-monotonic behavior of Hall resistivity. As domain wall resistance increases, the THE-like features become more pronounced.
Additionally, we highlighted the limitations of assuming a purely linear anomalous Hall effect in interpreting Hall measurements, as real systems often exhibit nonlinear contributions that complicate the analysis. \textcolor{black}{Due to the variation of the reciprocal space Berry phase under the field-driven magnetization evolution, the resulting anomalous Hall resistivity is sometimes no longer a linear function of the magnetization. Usually due to the drastic change of the Berry phase, the hump shows up in the raw data of the Hall resistivity, like in the case of EuCd$_2$As$_2$ where a giant nonlinear anomalous Hall effect has been observed \cite{Cao2022}. In contrast, our mechanism does not support a hump in the raw data. The hump shows up only when the linear anomalous Hall resistivity is excluded from the total Hall resistivity.}. Our findings have broad implications for studying Hall effects in disordered ferromagnets, thin films, and multilayer systems. As experimental techniques continue to improve, direct imaging of spin textures and complementary measurements will be essential to conclusively differentiate between topological and non-topological contributions to Hall resistivity. 


\section*{Acknowledgment}    
This work was supported by the Office of Basic Energy Sciences, Division of Materials Sciences and Engineering, U.S. Department of Energy, under Award No. DE-SC0020221. This research used resources of the National Energy Research Scientific Computing Center, a DOE Office of Science User Facility using NERSC award BES-ERCAP0030590.

\bibliographystyle{apsrev4-1}

\end{document}